\documentclass[11pt,twoside]{article}
\usepackage{asp2010}

\resetcounters

\begin{document}

\title{Astrophysics Source Code Library}

\author{Alice Allen$^1$, Kimberly DuPrie$^1$, Bruce Berriman$^{2,4}$, Robert J. Hanisch$^{3,4}$, Jessica Mink$^5$, Peter J. Teuben$^6$
\affil{$^1$Astrophysics Source Code Library}
\affil{$^2$Infrared Processing and Analysis Center, California Institute of Technology}
\affil{$^3$Space Telescope Science Institute}
\affil{$^4$Virtual Astronomical Observatory}
\affil{$^5$Harvard-Smithsonian Center for Astrophysics}
\affil{$^6$Astronomy Department, University of Maryland}
}

\begin{abstract}

The Astrophysics Source Code Library (ASCL), founded in 1999, is a
free on-line registry for source codes of interest to astronomers and
astrophysicists. The library is housed on the discussion forum for
Astronomy Picture of the Day (APOD) and can be accessed at
http://ascl.net. The ASCL has a comprehensive listing that covers a
significant number of the astrophysics source codes used to generate
results published in or submitted to refereed journals and continues
to grow. The ASCL currently has entries for over 500 codes; its
records are citable and are indexed by ADS. The editors of the ASCL
and members of its Advisory Committee were on hand at a demonstration table in the ADASS poster room to present
the ASCL, accept code submissions, show how the ASCL is starting to be
used by the astrophysics community, and take questions on and
suggestions for improving the resource.
\end{abstract}

\section{Introduction}

The Astrophysics Source Code Library is a free on-line registry of source codes used in research by astronomers and astrophysicists. With 546 codes indexed as of October 31, 2012, it is the largest resource for astrophysics software in existence. 

The ASCL takes an active approach to adding codes, seeking out new and old codes used in research published in or submitted to refereed journals. Housed on a phpbb board, the site is easy to use and offers full-text iterative searching. Its affiliation with APOD offers stable, ongoing exposure for the resource. Since January, 2012, the SAO/NASA Astrophysics Data System (ADS) has indexed ASCL entries, making codes in it both easily discoverable and citable.

\section{History of the ASCL}

The ASCL was created in 1999 and collected 37 codes by 2002. As other code libraries started up, updates to the ASCL ceased though the repository remained continuously available. In mid-2010, the ASCL moved to a new site under the direction of founder Robert Nemiroff and new editor Alice Allen, and a period of expansion of the registry started, with an average of 18 codes added to the resource every month since the move. In 2011, an Advisory Committee (AC) chaired by Peter J. Teuben (Astronomy Department, University of Maryland) was established and includes: 
\begin{itemize}
\item Bruce Berriman, Infrared Processing and Analysis Center, California Institute of Technology/Virtual Astronomical Observatory
\item Robert J. Hanisch, Space Telescope Science Institute/Virtual Astronomical Observatory
\item Jessica Mink, Harvard-Smithsonian Center for Astrophysics
\item Robert J. Nemiroff, Michigan Technological University
\item Lior Shamir, Lawrence Technological University
\item John Wallin, Middle Tennessee State University
\end{itemize}

\section{ASCL demonstration table}

A table was set up at ADASS and manned by the editors to provide information on and demonstrate the ASCL, answer questions, distribute the handout for the Monday afternoon BoF session {\em Bring out your codes!} and a list of relevant articles, provide updated information, and accept code submissions and suggestions for making the resource more useful. Peter J. Teuben,  Bruce Berriman, Robert J. Hanisch, and Jessica Mink, members of the AC, were available at different times throughout the conference to discuss the registry. Slide shows about the ASCL ran on the display monitor when the registry was not being demonstrated.

\subsection{Use of ASCL entries by the community}
The ASCL editors shared information with ADASS attendees on how ASCL entries are starting to be used in the community, including: 
\begin{itemize}

\item as citations in papers\footnote{http://arxiv.org/pdf/1207.5048v1} and documentation.\footnote{http://www2.astro.psu.edu/xray/docs/TARA/ae\_users\_guide/node21.html}

\item in pre-prints for the location of the software.\footnote{http://arxiv.org/abs/1204.0066}

\item in publications lists\footnote{http://www.nottingham.ac.uk/physics/people/adam.moss\#lookup-publications, http://abiliomateus.net/astro/pub/proceedings} and Google Scholar citation indices.\footnote{http://scholar.google.fr/citations?user=wBd3KzgAAAAJ\&hl=en}

\item to link papers to the software they use.\footnote{http://adsabs.harvard.edu/cgi-bin/bib\_query?arXiv:astro-ph/0402443}

\end{itemize}

\subsection{Update on growth and activity}
Though we are starting to see more codes submitted by their authors as more people learn of the ASCL, in the past two years nearly all codes have been added to the ASCL by the editors. Figure~\ref{growthbyquarter} shows the growth in the number of codes in the registry from the third quarter, 2010 through the third quarter, 2012.

\articlefigure{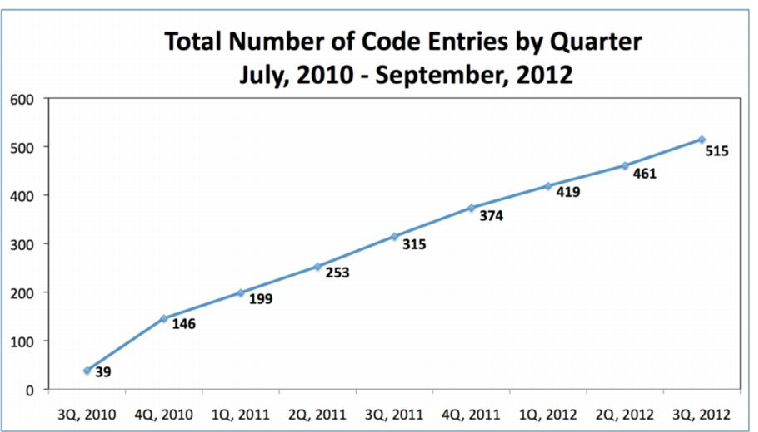}{growthbyquarter}
{Number of entries in the ASCL since mid-2010}

Activity on the ASCL is influenced by several factors, including the release of blog posts and articles about the ASCL, presentations at conferences, and links from APOD to the forum in which the resource resides. Google Analytics was installed in 2011 to measure activity on the site; pageviews by quarter are shown in Figure~\ref{pageviewsbyquarter}. Though activity varies, it is trending upward over time. 

\articlefigure{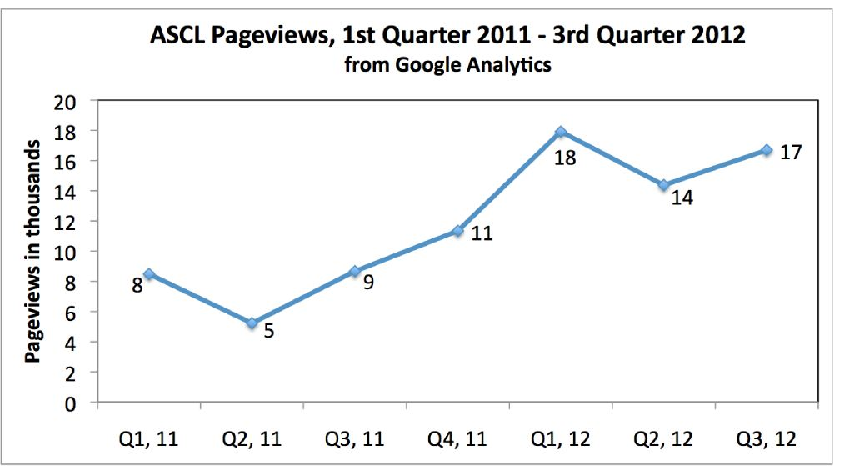}{pageviewsbyquarter}
{Pageviews in thousands as captured by Google Analytics}

\section{Resources available online for download or viewing}
The presentations displayed at the demonstration table are available either on the ASCL blog\footnote{http://asterisk.apod.com/wp/?page\_id=61} or as PowerPoint downloads.\footnote{http://asterisk.apod.com/wp/?p=216} The {\em Bring out your codes!} talking points\footnote{http://asterisk.apod.com/library/ASCL/ADASS2012/BoFTalkingPoints.docx} and the list of relevant articles\footnote{http://asterisk.apod.com/library/ASCL/ADASS2012/Papers\%20of\%20Possible\%20Interest\%20to\%20Astronomical\%20Software\%20Users.docx} can be downloaded from the ASCL blog, as can a flyer about the ASCL.\footnote{http://asterisk.apod.com/wp/?p=216}

\acknowledgements The ASCL thanks the American Astronomical Society for its support.

\bibliography{D1}

\end{document}